\documentstyle[12pt,epsf,amsfonts,amssymb,cite,amsbsy]{article}
\newcommand{\Rsub}{\rm\scriptscriptstyle}

\def\slashchar#1{\setbox0=\hbox{$#1$}           
   \dimen0=\wd0                                 
   \setbox1=\hbox{/} \dimen1=\wd1               
   \ifdim\dimen0>\dimen1                        
      \rlap{\hbox to \dimen0{\hfil/\hfil}}      
      #1                                        
   \else                                        
      \rlap{\hbox to \dimen1{\hfil$#1$\hfil}}   
      /                                         
   \fi}                                         %
\thicklines

\begin{document}

\begin{center}
{\large\bf Leptonic constants of heavy quarkonia\\[2mm] in potential approach
of NRQCD}\\[7mm]
    \renewcommand\thefootnote{\star}%
{\sf V.V.Kiselev}\footnote{E-mail: kiselev@th1.ihep.su}, {\sf A.K.Likhoded},
\\
\vspace*{3mm}
Russian State Research Center ``Institute for High Energy Physics",\\
Protvino, Moscow Region, 142284 Russia\\
\vspace*{3mm}
and\\
\vspace*{3mm}
{\sf O.N.Pakhomova}, {\sf V.A.Saleev},
\\
\vspace*{3mm}
Samara State University,\\
Samara, Pavlov 1, 443011 Russia
\end{center}

\vspace*{2mm}
\begin{abstract}
We consider a general scheme for calculating the leptonic constant of heavy
quarkonium $\bar Q Q$ in the framework of nonrelativistic quantum
chromodynamics, NRQCD, operating as the effective theory of nonrelativistic
heavy quarks. We explore the approach of static potential in QCD, which takes
into account both the evolution of effective charge in the three-loop
approximation and the linearly raising potential term, which provides the quark
confinement. The leptonic constants of $\bar b b $ and $\bar c c$ systems are
evaluated by making use of two-loop anomalous dimension for the current of
nonrelativistic quarks, where the factor for the normalization of matrix
element is introduced in order to preserve the renormalization group invariance
of estimates.
\end{abstract}
\setcounter{footnote}{0}
    \renewcommand\thefootnote{\arabic{footnote}}%

\vspace*{1cm}

\vspace*{1cm} 
\section{Introduction}
A description of leptonic constants for a heavy quarkonium composed of heavy
quark and heavy anti-quark $\bar Q Q$ plays an important role in studies of QCD
dynamics in the sector of heavy quarks. So, the precision calculations of heavy
quark masses $m_{b,c}$ and running coupling constant $\alpha_s$ in QCD are
performed in the framework of QCD sum rules \cite{SVZ} and effective theory of
nonrelativistic heavy quarks, NRQCD \cite{NRQCD}. In such the approach the
experimental data on both the masses of $S$-wave quarkonium states and their
leptonic constants \cite{PDG} are used. The consideration of NRQCD sum rules
was done for the bottomonium with taking into account of one-loop gluon
corrections in \cite{mbv} (next-to-leading-order, NLO) as well as with two-loop
corrections in \cite{pp,ben,hoang} (next-to-next-to-leading-order, N$^2$LO),
and for the charmonium in \cite{eide}. In general, the sum rules of NRQCD could
be used for the solution of inverse problem, i.e. the calculation of leptonic
constants at given values of heavy quark masses, a definite normalization of
coupling constant and the experimental mass spectrum of quarkonia. The only
additional entry is caused by that the analysis in the framework of NRQCD sum
rules supposes that the region under study remains correctly defined only if
one takes into account the higher excitations of ground state, so that  one
should fix the ratios of leptonic constants, i.e. their relative weights in the
sum rules, along with the mass spectrum. Another way is to go in the region of
parameters, where the contributions of ground state excitations are negligible,
and the term of gluon condensate becomes significant, while its Wilson
coefficient requires cumbersome calculations in higher orders of perturbation
theory.

An alternative method is the description of heavy quarkonium in the framework
of potential approach. In this way, if the static potential in QCD with
infinitely heavy sources in the fundamental representation of color gauge group
is given, then in the leading order of NRQCD, i.e., if we neglect relativistic
corrections and interactions depending on the quark spins, the heavy quark
masses can be determined with a high accuracy from the comparison of
experimental mass spectrum of heavy quarkonia $\bar b b$ and $\bar c c$ with
the estimates in the NRQCD. In this procedure, one calculates the spectra of
stationary energy levels $E$ for the nonrelativistic Schr\"odinger equation
with the static potential, so that the masses are defined by the expression $M
= m_1+m_2+E$, where $m_{1,2}$ are the masses of heavy quark and heavy
anti-quark composing the quarkonium. In addition, in ref. \cite{3loop} authors
derived the static potential in QCD, so that it takes into account recent
perturbative results at short distances \cite{Peter,Schroed} as well as the
regime of confinement in the infrared region. This potential is straightforward
modification of model by Buchm\"uller and Tye \cite{BT}, who suggested the
method with unified two-loop $\beta$-function for the effective charge
${\mathfrak
a}$. This function satisfied two definite asymptotic limits at ${\mathfrak
a}\to
0$ (the perturbation theory) and ${\mathfrak a}\to \infty$ (the potential
confining the quarks in hadrons, in the form of linearly raising term under the
distance increase). The calculations in three loops \cite{Peter,Schroed} shown
that the value and sign of coefficient $\beta_2$ modelled in the paper by
Buchm\"uller and Tye are significantly different from the correct results for
the $\beta$-function. The modification allows us to reach a consistent
description for combined data on the mass spectra of heavy quarkonia and the
evolution parameter $\Lambda_{QCD}$ for the running coupling constant measured
at high virtuality of $m_Z$ \cite{PDG}. Thus, in the potential approach we have
got a possibility to estimate the leptonic constants of heavy quarkonia in
NRQCD with the quark masses and coupling constant $\alpha_s$ adjusted with a
high accuracy.

In this way, we consider a general scheme for the calculation of leptonic
constants, since the results of NRQCD have to be related with the quantities
defined in full QCD. There are two challenges in this respect. The first one is
the Wilson coefficient depending on the scale of calculations in NRQCD, which
relates the current of nonrelativistic quarks considered in the leading order
of inverse heavy quark mass, with the electromagnetic current of heavy quarks
in full QCD. An anomalous dimension of this coefficient ${\cal K}$ is known
with the two-loop accuracy \cite{HT,bensmir,melch}. The factor ${\cal K}$ is
matched with the value, which is given by the equality of currents in QCD and
its nonrelativistic approximation at a scale $\mu_{\rm hard}$ about the mass of
heavy quark. Therefore, the operator equality of currents in full QCD and NRQCD
is completely defined up to the two-loop accuracy. The second challenge is the
matrix element for the current of nonrelativistic quarks over the vacuum and
physical state of quarkonium. Because of renormalization group invariance of
physical quantities, this matrix element has got the anomalous dimension equal
to that of the current in NRQCD. For the first time, in the potential approach
we perform the consideration involving the factor ${\cal A}$ providing the
renormalization invariance of results. In this way, we have to introduce a
scale of normalization for ${\cal A}$, so that the reasonable region of scale
change is restricted by the physical characteristics for the quark system under
study. Then, the obtained estimates of leptonic constants for the quarkonia are
stable in a region of change for both the QCD-to-NRQCD matching scale $\mu_{\rm
hard}$ and the point of perturbative calculations in NRQCD $\mu_{\rm fact}$.

In Section 2 we consider the description of leptonic constants for the heavy
quarkonia in the framework of NRQCD. The potential approach is actually given
in Section 3. The numerical estimates are presented in Section 4. In
Conclusion we summarize the obtained results.

\section{Leptonic constants}

In the NRQCD approximation for the heavy quarks, the calculation of leptonic
constant for the heavy quarkonium with the two-loop accuracy requires the
matching of NRQCD currents with the currents in full QCD,
$$
J_\nu^{QCD}= \bar Q \gamma_\nu Q, \;\;\; {\cal J}_\nu^{NRQCD} = \chi^\dagger
\sigma_\nu^\perp \phi,
$$
where we have introduced the following notations: $Q$ is the relativistic quark
field, $\chi$ and $\phi$ are the nonrelativistic spinors of anti-quark and
quark, $\sigma_\nu^\perp= \sigma_\nu -v_\nu (\sigma \cdot v)$, where $v$ is the
four-velocity of heavy quarkonium, so that
\begin{equation}
J_\nu^{QCD} = {\cal K}(\mu_{\rm hard}; \mu_{\rm fact})\cdot {\cal
J}_\nu^{NRQCD}(\mu_{\rm fact}),
\label{match}
\end{equation}
where the scale $\mu_{\rm hard}$ gives the normalization point for the matching
of NRQCD with full QCD, while $\mu_{\rm fact}$ denotes the normalization point
for the calculations in the perturbation theory of NRQCD. 

For the heavy quarkonium composed of heavy quarks with the same flavor, the
Wilson coefficient ${\cal K}$ is known with the two-loop accuracy
\cite{HT,bensmir,melch,mel}
\begin{equation}
{\cal K}(\mu_{\rm hard}; \mu_{\rm fact}) = 1 -\frac{8}{3}
\frac{\alpha_s^{\overline{\Rsub MS}}(\mu_{\rm hard})}{\pi}+
\left(\frac{\alpha_s^{\overline{\Rsub MS}}(\mu_{\rm
hard})}{\pi}\right)^2 c_2(\mu_{\rm hard}; \mu_{\rm fact}),
\label{kfact}
\end{equation}
and $c_2$ is explicitly given in \cite{bensmir,melch}. The anomalous dimension
of factor ${\cal K}(\mu_{\rm hard}; \mu_{\rm fact})$ in NRQCD is defined by 
\begin{equation}
\frac{d \ln{\cal K}(\mu_{\rm hard}; \mu)}{d \ln \mu} = \sum_{k=1}^{\infty}
\gamma_{[k]} \left(\frac{\alpha_s^{\overline{\Rsub MS}}(\mu)}{4\pi}\right)^k,
\label{anom}
\end{equation}
whereas the two-loop calculations\footnote{We use ordinary notations for the
invariants of $SU(N_c)$ representations: $C_F=\frac{N_c^2-1}{2 N_c}$, $C_A=
N_c$, $T_F = \frac{1}{2}$, $n_f$ is a number of ``active'' light quark
flavors.} give
\begin{eqnarray}
\gamma_{[1]} & = & 0,\\
\gamma_{[2]} & = & -16 \pi^2 C_F \left(\frac{1}{3} C_F + \frac{1}{2}
C_A\right). \label{g2}
\end{eqnarray}
The initial condition for the evolution of factor ${\cal K}(\mu_{\rm hard};
\mu_{\rm fact})$ is given by the matching of NRQCD current with full QCD at
$\mu = \mu_{\rm hard}$ \cite{bensmir,melch}.

The leptonic constant of heavy quarkonium is defined in the following way:
\begin{equation}
\langle 0| J_\nu^{QCD} |\bar Q Q ,\lambda \rangle = \epsilon_\nu^\lambda
f_{\bar Q Q } M_{\bar Q Q },
\end{equation}
where $\lambda$ denotes the vector state polarization $\epsilon_\nu$. In full
QCD the electromagnetic current of quarks is conserved, while in
NRQCD the current ${\cal J}_\nu^{NRQCD}$ has the nonzero anomalous dimension,
so that in accordance with (\ref{match})--(\ref{g2}), 
we find 
\begin{equation}
\langle 0| {\cal J}_\nu^{NRQCD}(\mu) |\bar Q Q ,\lambda \rangle = {\cal
A}(\mu)\; \epsilon_\nu^\lambda f_{\bar Q Q }^{NRQCD} M_{\bar Q Q },
\label{a}
\end{equation}
where, in terms of nonrelativistic quarks, the leptonic constant for the heavy
quarkonium is given by the well-known relation with the wave function at the
origin 
\begin{equation}
f_{\bar Q Q}^{NRQCD} = \sqrt{\frac{12}{M}}\; |\Psi_{\bar Q Q}(0)|,
\label{wave}
\end{equation}
and the value of wave function in the leading order is determined by the
solution of Schr\"odinger equation with the static potential, so that we
isolate the scale dependence of NRQCD current in the factor ${\cal A(\mu)}$,
while the leptonic constant $f_{\bar Q Q}^{NRQCD}$ is evaluated at a fixed
normalization point $\mu=\mu_0$, which will be attributed below. It is evident
that
\begin{equation}
f_{\bar QQ} = f_{\bar QQ}^{NRQCD} {\cal A}(\mu_{\rm fact})\cdot {\cal
K}(\mu_{\rm hard}; \mu_{\rm fact}),
\label{cc}
\end{equation}
and the anomalous dimension of ${\cal A}(\mu_{\rm fact})$ should compensate the
anomalous dimension of factor ${\cal K}(\mu_{\rm hard}; \mu_{\rm fact})$, so
that in two loops we have got 
\begin{equation}
\frac{d \ln{\cal A}(\mu)}{d \ln \mu} = - \gamma_{[2]}
\left(\frac{\alpha_s^{\overline{\Rsub MS}}(\mu)}{4\pi}\right)^2.
\label{anoma}
\end{equation}
The physical meaning of ${\cal A}(\mu)$ is clearly determined by the relations
of (\ref{a}) and (\ref{cc}): this factor gives the normalization of matrix
element for the current of nonrelativistic quarks expressed in terms of wave
function for the two-particle quark state (in the leading order of inverse
heavy quark mass in NRQCD). Certainly, in this approach the current of
nonrelativistic quarks is factorized from the quark-gluon sea, which is a
necessary attribute of hadronic state, so that, in general, this physical state
can be only approximately represented as the two-quark bound state. In the
consideration of leptonic constants in the framework of NRQCD, this
approximation requires to introduce the normalization factor ${\cal A}(\mu)$
depending on the scale.

The renormalization group equation of (\ref{anoma}) is simply integrated out,
so that
\begin{equation}
{\cal A}(\mu) = {\cal A}(\mu_0)\; \left[ \frac{\beta_0+\beta_1
{\displaystyle\frac{\alpha_s^{\overline{\Rsub
MS}}(\mu)}{4\pi}}}{\beta_0+\beta_1 {\displaystyle
\frac{\alpha_s^{\overline{\Rsub MS}}(\mu_0)}{4\pi}}}
\right]^{\displaystyle\frac{\gamma_{[2]}}{2\beta_1}},
\end{equation}
where $\beta_0 = \frac{11}{3} C_A - \frac{4}{3} T_F n_f$, and
$\beta_1 = \frac{34}{3} C_A^2 - 4 C_F T_F n_f  - \frac{20}{3} C_A T_F n_f$.
A constant of integration could be defined so that at a scale $\mu_0$ we would
get ${\cal A}(\mu_0)=1$. Thus, in the framework of NRQCD we have got the
parametric dependence of leptonic constant estimates on the scale $\mu_0$,
which has the following simple interpretation: the normalization of matrix
element for the current of nonrelativistic quarks at $\mu_0$ is completely
given by the wave function of two-quark bound state. At other $\mu\ne \mu_0$ we
have to introduce the factor ${\cal A}(\mu)\ne 1$, so that the approximation of
hadronic state by the two-quark one becomes inexact.

The relations above are general for the NRQCD description of leptonic constant
for the heavy quarkonium in both the approach of sum rules and the calculations
with the static potential, so that the only difference is the use of
(\ref{wave}) in the potential approach. In the sum rules this relation is
substituted by the equations for the constants $f_{\bar QQ}^{NRQCD}$, which
follow from the quark-hadron duality in the corresponding scheme of
calculations. In this respect, we have to stress that 
the analysis of NRQCD sum rules for the bottomonium in the two-loop
approximation in \cite{pp,hoang,mel} took into account the factor of ${\cal
A}(\mu)$ in a different way. So, in the calculation of two-point correlator for
the currents of nonrelativistic quarks the regularization is necessary in the
two-loop order. Of course, the anomalous dimension of two-point current
correlator compensates the anomalous dimension of corresponding Wilson
coefficient. In the sum rules with the one-loop gluon correction \cite{mbv}
this factor can be omitted, since its one-loop anomalous dimension is equal to
zero, and, hence, the factor can be supposed equal to unit. As we have shown
taking into account the two-loop corrections, the calculations without the
normalization factor ${\cal A}(\mu)$ for the matrix element of NRQCD current
would be, in general, not correct.

Some comments are to the point. It is worth to note that we deal with the
operator product expansion (OPE), wherein the electromagnetic current of heavy
quarks is expressed in terms of leading $1/m_Q$-order current of
nonrelativistic spinors near the threshold. Therefore, in this OPE the
corresponding Wilson coefficient, i.e. the matching factor ${\cal K}$, appears.
The first problem is a calculation of this factor. The second is an evaluation
of matrix element for the NRQCD current of nonrelativistic quarks. These two
problems should be clearly distinguished in the estimates. The calculation of
Wislon coefficient ${\cal K}$ can be done \underline{perturbatively} in two
ways. Originally, it was done in the technique of threshold expansion of loop
integrals for Feynman diagrams in full QCD for heavy quarks
\cite{HT,bensmir,melch,mel}. We use this result throughout this paper. Further,
the second way of ${\cal K}$ calculation is the following: one should
perturbatively expand the QCD Lagrangian for the interacting heavy quark and
heavy anti-quark separated by a distance $r$ in terms of nonrelativistic
spinors and represent the perturbative potential contributions different from
the interactions with soft and ultrasoft fields in the hadron. These soft terms
are not essential in the perturbative calculations of matching factor ${\cal
K}$ at high virtualities, and under such conditions the system of heavy quark
and heavy anti-quark can be considered as the coulomb system in the leading
approximation, while the potential corrections such as the spin-dependent and
relativistic terms in the form of $1/m_Q$ contributions converted to the
$\alpha_s$ corrections to the energy, since in the coulomb systems the distance
behaves as $r\sim 1/\alpha_s m_Q\sim 1/|{\bf p}|$. Anyway, the main point is
that the factor ${\cal K}$ can be calculated in the {\it perturbative theory}.
It is important that the fact of cancellation of anomalous dimension for the
factor ${\cal K}$ by the anomalous dimension for the NRQCD current follows from
the zero anomalous dimension of electromagnetic current in full QCD with no
reference to the calculation procedure, of course. Thus, the cancellation has
no connection to the problem of evaluation for the hadronic matrix element of
NRQCD current. Indeed, the effective expansion of QCD Lagrangian for the heavy
quarks in the form of potential nonrelativistic QCD (pNRQCD) \cite{pNRQCD} or
velocity-counting nonrelativistic QCD (vNRQCD, discussed below)
\cite{LMR,MS,HMS} cannot be applied to such the calculation of matrix element
because the leading coulomb approximation does not dominate in comparison with
corrections for the real quarkonia, even for the $\bar b b(1S)$. So, in pNRQCD
and vNRQCD the potentials are treated as Wilson coefficients in front of four
quark operators, while the soft and ultrasoft terms are separated. This
separation is valid only if the time for the formation of coulomb wave function
is many times less than the time of interaction with the soft fields (a
quark-gluon sea and a string). Then,
$$
T_{coul} \ll T_{soft}.
$$
These times can be easily evaluated in terms of distance between the quarks,
their velocity and soft scale $\Lambda_{QCD}$, so that
$$
\left.
\begin{tabular}{r}
$
\displaystyle
T_{coul} \sim \frac{r}{v}\sim \frac{1}{m_Q v^2}
$\\[4mm]
$
\displaystyle
T_{soft} \sim \frac{1}{\Lambda_{QCD}}
$
\end{tabular}
\right\} \Longrightarrow \;\;\; m_Q v^2 \gg \Lambda_{QCD}.
$$
Therefore, in order to allow one to correctly use the pNRQCD or vNRQCD to the
description of matrix elements, the kinetic energy of heavy quark motion should
be many times greater than the scale of confinement.

In practice, estimates for the most deep `coulomb' system $b\bar b(1S)$ give
$$
m_Q v^2 \sim 500 \mbox{ MeV,}\;\;\;\;\mbox{ at }  \Lambda_{QCD}\sim 300 \mbox{
MeV.} 
$$
Thus, in order to get the potential model applicable to the real quarkonia, one
should include {\it soft terms} in the {\it static} energy, i.e. the
{\it static potential}. This static potential has an infrared stability and it
is independent of any separation of soft and non-soft terms in the static
energy. The static potential defined in terms of Wilson loop possesses these
properties and it does not coincide with the potential (Wilson coefficient) in
pNRQCD or vNRQCD, since the infrared contributions are added, but separated.
Therefore, we deal with the model, wherein the total energy of all dynamical
fields in the presence of static sources is taken into account. By the way, the
relativistic and spin-dependent corrections in this case determine the scale of
corrections about 50 MeV and their contribution gives the uncertainty of
estimation, which is many times less than the uncertainty due to the scale
variation in the estimates for the leptonic constants with the static
potential.

We stress that there is an ordinary situation in the leptonic constant
calculation, since the perturbative technique used to evaluate the Wislon
coefficient cannot be explored in order to get a correct estimate for the
hadronic matrix element for the operator in front of the Wislon coefficient.

Recently, the effective theory for the heavy nonrelativistic quark and
antiquark pair in QCD was developed on the basis of relative velocity
expansion \cite{LMR,MS,HMS}. This approach uses the so-called velocity
renormalization group, so that we refer to such the theory as vNRQCD. The
physical argumentation of vNRQCD is the following: In the heavy quark-antiquark
system one expects a low influence of light quark-gluon sea formed by
nonperturbative emission of quarks and gluons with virtualities about
$\Lambda_{QCD}$, so that the heavy quark motion is close to the coulomb
approximation up to higher order corrections in $\alpha_s$. Moreover, we can
suppose the coulomb relations between various scales, which possess the
following counting rules: the relative velocity $v\sim \alpha_s$, the momentum
transfer (or the inverse distance between the quarks), i.e. the soft scale
$\mu_S \sim |\boldsymbol k|\sim m_Q\cdot v$, the momentum of heavy quark
$|\boldsymbol p|\sim \mu_S$, while the kinetic and potential energies, i.e. the
ultrasoft scale, $\mu_U \sim E\sim V\sim m_Q\cdot v^2$, so that in the
perturbative part of effective Lagrangian the distance, momentum and energy are
not free independent quantities, since they are related by the single
``running'' parameter $v$. This approach is certainly different from pNRQCD,
wherein the static bilinear operators of nonrelativistic spinors are fixed at a
short distance, while the kinetic energy as well as the quark momenta are put
to zero.

Thus, the starting point of vNRQCD is the correlation of soft and ultrasoft
scales, $\mu_U\approx v\cdot \mu_S$, in contrast to pNRQCD. Therefore, in
vNRQCD the matching with the full QCD at the scale of heavy quark mass
$\mu_{\rm hard} = m_Q$ and the renormalization group evolution with $v$
involving large $\ln v$ result in a combined summation\footnote{Note that terms
$[\ln v]^n \sim [\ln \alpha_s]^n$ reproduce so-called manifestly nonanalytic
contributions in $\alpha_s$-expansion.} of relevant momentum and energy
contributions from $\mu =m_Q$ up to $\mu = m_Q\cdot v^2$, while pNRQCD deals
with the two-step matching and evolution: NRQCD matched with full QCD at
$\mu_{\rm hard} \sim m_Q$ and evolved to $\mu_S\sim m_Q\cdot v$, and pNRQCD
matched with NRQCD at $\mu_S$ and evolved to $\mu\sim m_Q\cdot v^2$. A peculiar
feature of vNRQCD summing up both logs of $\mu_S$ and $\mu_U$ is that the
limits of $m_Q\to\infty$ or $v\to 0$ do not reproduce the static limit (see
explanations provided by examples in \cite{HMS}). 

Next point is the anomalous dimension of nonrelativistic current, that was
calculated in \cite{LMR,MS} in the technique of vNRQCD. One found that
expressions (\ref{anom})-(\ref{g2}) are reproduced in vNRQCD and strictly valid
at $\mu=m_Q$, while at lower virtualities with $v = \mu_{\rm fact}/m_Q < 1$
some corrections in $\ln v$ appear. These corrections can be summed up in
vNRQCD, so that substitution of $\alpha_s^2(\mu)$ for $\alpha_s^2(m_Q)$ in the
two-loop anomalous dimension could be quite naive and not strictly justified,
by opinion in \cite{MS}. The analogous two-loop anomalous dimension $\gamma$ of
$\cal K$ in vNRQCD as well as the complete expression for ${\cal K}^{\rm
vNRQCD}(v)$ integrated out in the velocity renormalization group are available
in \cite{MS}. So, 
\begin{equation}
{\cal K}^{\rm vNRQCD}(v) = {\cal K}^{\rm vNRQCD}(1)\,\exp\left[ \sum
\limits_{i=1}^{7} a_i^v \pi \alpha_s(m_Q)k_i\right],
\end{equation}
where $a_i^v$ are expressed in terms of $C_F$, $C_A$ and $\beta_0$, and the
functions $k_i$ have the form
$$
\begin{array}{lll}
\displaystyle k_1 = \frac{1}{z}-1, & k_2 = 1-z,\;\; & k_3 = \ln(z),\\[4mm]
k_4 = 1- z^{1-13C_A/(6\beta_0)},\;\; & \multicolumn{2}{l}{k_5 =
1- z^{1-2C_A/\beta_0},} \\[4mm]
\multicolumn{3}{l}{\displaystyle k_6 = \frac{\pi^2}{12}-\frac{1}{2}
\ln^2(2)-\ln(w) \ln \left(\frac{2w}{2w-1}\right) -{\rm
Li}_2\left(\frac{1}{2w}\right),} \\[4mm]
\multicolumn{3}{l}{\displaystyle k_7 = \frac{w}{2w-1}\ln(w)
-\frac{1}{2}\ln(2w-1),}
\end{array}
$$
while $z = \alpha_s(\mu)/\alpha_s(m_Q)$, $w= \alpha_s(\mu^2/m_Q)/
\alpha_s(\mu)$, and the one-loop matching condition is given by the same form
as in (2), i.e.
\begin{equation}
{\cal K}^{\rm vNRQCD}(1) = 1- 2C_F\, \frac{\alpha_s(m_Q)}{\pi}.
\end{equation}
However, in general the scale prescription in the argument of $\alpha_s$
standing in the first nonzero term, which presents the accuracy under study,
is rather conventional, since the scale change results in the next-order
contribution in $\alpha_s$, that is not under control. Indeed, the substitution
$$
\alpha_s(\mu) \approx \alpha_s(m_Q) \left(1 -
\frac{\beta_0}{2\pi}\,\alpha_s(m_Q)\,\ln\frac{\mu}{m_Q}\right) = 
\alpha_s(m_Q) \left(1 - \frac{\beta_0}{2\pi}\,\alpha_s(m_Q)\,\ln v\right),
$$
influences the three-loop anomalous dimension $\gamma$ for $\cal K$, only. The
argumentation of vNRQCD is that the corrections summing up $[\ln v]^n$ are
dominant (in all orders in $\alpha_s$). So, the above note on the scale
substitution manifests itself in three-loop running of coulomb-like potentials
in vNRQCD.

In the present paper we follow the static ideology with the threshold
expression for ${\cal K}(\mu)$ (and hence, ${\cal A}(\mu)$), since we adopt the
Schr\"odinger equation with the model of static potential for the $\bar Q Q$
system in contrast to the coulomb approximation improved by the velocity
renormalization group involving running coulomb and relevant potentials.
Moreover, the threshold expression for ${\cal K}(\mu)$ is preferable by
technical reasons, since, first, it is valid at the two-loop order (as we have
mentioned above), second, ${\cal K}(\mu)$ has an explicit dependence on the
matching scale $\mu_{\rm hard}$, that allows us to study the point stable under
the variation of $\mu_{\rm hard}$ (see below), while in vNRQCD the matching
scale is fixed and prescribed to $\mu_{\rm hard} = m_Q$, and the corresponding
analysis on the stability is not possible. Third, the factor ${\cal K}(\mu)$
contains the two-loop final term independent of $\ln \mu$, while ${\cal K}^{\rm
vNRQCD}(v)$ does not (except the one-loop term), since in the large $\ln v$
approximation the two-loop final term is irrelevant. This two-loop final term
in ${\cal K}(\mu)$ makes the cancellation of $\mu$-dependence in the product of
${\cal K}(\mu)\cdot {\cal A}(\mu)$ valid in a restricted region of $\mu$, that
is conceptually true, while in vNRQCD the large log approximation results in
the complete cancellation of $v$-dependence (remember, $v = \mu_{\rm
fact}/m_Q$) in the product of ${\cal K}^{\rm vNRQCD}(\mu)\cdot{\cal A}^{\rm
vNRQCD}(\mu)$. Nevertheless, since we assume the existence of $\mu_0$, at which
${\cal A}(\mu_0)=1$, then we can estimate the above product by ${\cal K}^{\rm
vNRQCD}(v_0)$. The problem on the choice of $\mu_0$ in vNRQCD is more
artificial than in NRQCD, since the analysis of stability on $\mu_{\rm hard}$
results in a preferable prescription for $\mu_0$ in NRQCD, while in vNRQCD we
have no analogous criterion (see Section 4 with numerical results).

\section{Potential approach}

\begin{figure}[th]
\setlength{\unitlength}{1mm}
\begin{picture}(70,110)
\put(30,0){\epsfxsize=11cm \epsfbox{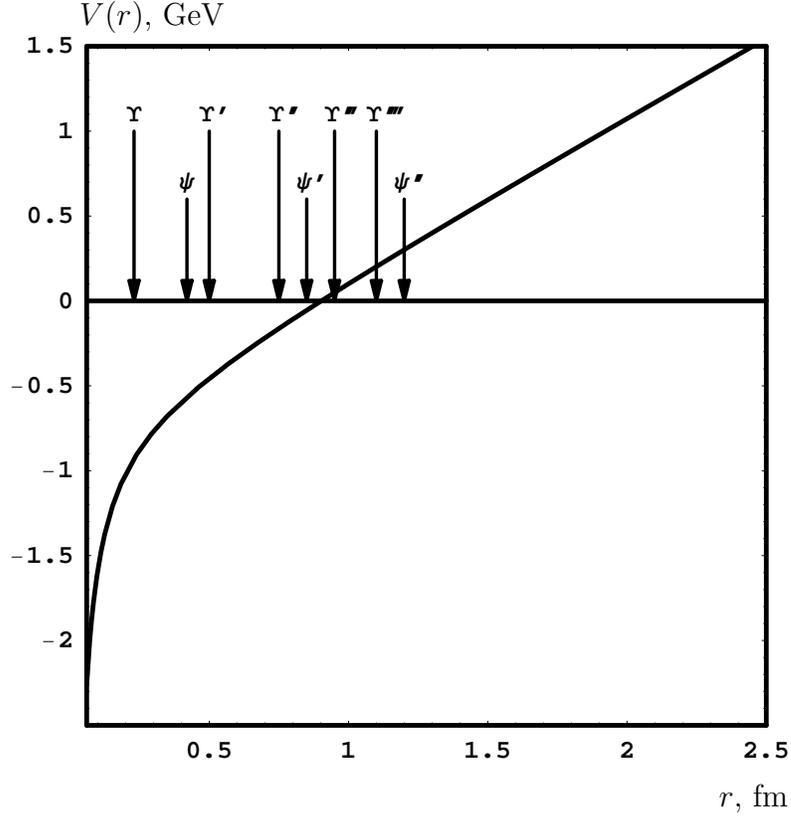}}
\put(130,3){$r$, fm}
\put(45,107){$V(r)$, GeV}
\end{picture}
\caption{The Cornell model of static potential and sizes of observed heavy
quarkonia with charmed quarks (the family $\psi$) and bottom quarks (the family 
$\Upsilon$).}
\label{gre}
\end{figure}

The potential of static heavy quarks illuminates the most important features of
QCD dynamics: the asymptotic freedom and confinement. In the leading order of
perturbative QCD at short distances and with a linear confining term in the
infrared region, the potential of static heavy quarks was considered in the
Cornell model \cite{Corn}, incorporating the simple superposition of both
asymptotic limits (the effective coulomb and string-like interactions). The
observed heavy quarkonia posed in the intermediate distances, where both terms
are important for the determination of mass spectra (see Fig. \ref{gre}). So,
the phenomenological approximations of potential (logarithmic one \cite{log}
and power law \cite{Mart}), taking into account the regularities of such the
spectra, were quite successful \cite{RQ}. 

The quantities more sensitive to the global properties of potential are the
wave functions at the origin as related to the leptonic constants and
production rates. So, the potentials consistent with the asymptotic freedom to
one and two loops as well as the linear confinement were proposed by Richardson
\cite{Richard}, Buchm\"uller and Tye \cite{BT}, respectively.

In QCD the static potential is defined in a manifestly gauge invariant way by
means of the vacuum expectation value of a Wilson loop \cite{Su},
\begin{eqnarray}
\label{def_WL}
V(r) &=& - \lim_{T\rightarrow\infty} 
\frac{1}{iT}\, \ln \langle{\cal W}_\Gamma\rangle \;, \nonumber\\
{\cal W}_\Gamma &=& \widetilde{\rm tr}\, 
{\cal P} \exp\left(ig \oint_\Gamma dx_\mu A^\mu\right) \;.
\end{eqnarray}
Here, $\Gamma$ is taken as a rectangular loop with time extension $T$ and
spatial extension $r$. The gauge fields $A_\mu$ are path-ordered along the
loop, while the color trace is normalized according to $\widetilde{\rm
tr}(..)={\rm tr}(..)/{\rm tr}1\!\!1\,$. This definition corresponds to the
calculation of effective action for the case of two external sources fixed at a
distance $r$ during an infinitely long time period $T$, so that the
time-ordering coincides with the path-ordering. Moreover, the contribution into
the effective action by the path parts, where the charges have been separated
to the finite distance during a finite time, can be neglected in comparison
with the infinitely growing term of $V(r)\cdot T$. Let us emphasize that the
defined static potential is, by construction, the renormalization invariant
quantity, since the action, by definition, doe not depend on the normalization
point.

Generally, one introduces the  V scheme of QCD coupling constant by the
definition of QCD potential of static quarks in momentum space as follows:
\begin{equation}
 V({\bf q}^2)  =  -C_F\frac{4\pi\alpha_{\Rsub V}({\bf q}^2)}{{\bf q}^2},
\end{equation}
so that for the such-way introduced value of $\alpha_{\Rsub V}$ one can derive
some results at large virtualities in the perturbative QCD as well as at low
transfer momenta in the approximation of linear term in the potential confining
the quarks.

In this section, first, we discuss two regimes for the QCD forces between the
static heavy quarks: the asymptotic freedom and confinement. Then we follow the
method by Buchm\"uller and Tye and formulate how these regimes can be combined
in a unified $\beta$ function for $\alpha_{\Rsub V}$ obeyed both limits of
small and large QCD couplings.

In the perturbative QCD up to the two-loop accuracy, the quantity
$\alpha_{\Rsub V}$ can be matched with $\alpha_{\overline{\Rsub MS}}$ by the
relation
\begin{eqnarray}
\alpha_{\Rsub V}({\bf q}^2)  =  \alpha_{\overline{\Rsub MS}}(\mu^2)
\sum_{n=0}^2 \tilde{a}_n(\mu^2/{\bf q^2})
\left(\frac{\alpha_{\overline{\Rsub MS}}(\mu^2)}{4\pi}\right)^n
= 
\alpha_{\overline{\Rsub MS}}({\bf q}^2)
\sum_{n=0}^2 a_n\left(\frac{\alpha_{\overline{\Rsub MS}}({\bf q}^2)}
{4\pi}\right)^n. 
\label{vms}
\end{eqnarray}
At present, in (\ref{vms}) the coefficients of tree approximation
$a_0$, one-loop contribution $a_1$ and new two-loop results $a_2$ (see
\cite{Peter,Schroed}) are known. Note that expansion (\ref{vms}) 
cannot be straightforwardly extended to higher orders of perturbative QCD
because of infrared problems, that result in nonanalytic terms in the
three-loop perturbative potential as was first discussed by Appelquist, Dine
and Muzinich \cite{Su}.

After the introduction of ${\mathfrak a} =\frac{\alpha}{4 \pi}$, the $\beta$
function is actually defined by 
\begin{equation}
\frac{d {\mathfrak a}(\mu^2)}{d\ln\mu^2} = \beta({\mathfrak a})
  = - \sum_{n=0}^\infty \beta_n \cdot {\mathfrak a}^{n+2}(\mu^2),
\end{equation}
so that $\beta_{0,1}^{\Rsub V}=\beta_{0,1}^{\overline{\Rsub MS}}$ and
$\beta_2^{\Rsub V} = \beta_2^{\overline{\Rsub MS}}-a_1\beta_1^{\overline{\Rsub
MS}} + (a_2-a_1^2)\beta_0^{\overline{\Rsub MS}}$. 

The Fourier transform results in the position-space potential \cite{Peter}, so
that the perturbative potential, by construction, is independent of
normalization point, i.e. it is the renormalization group invariant. However,
in the problem under consideration the truncation of perturbative expansion,
wherein the coefficients do not decrease\footnote{Moreover, according to the
investigations of renormalon, the coefficients in the series of perturbation
theory for the potential increase in the factorial power, so that the series
has a meaning of asymptotic one.}, leads to a strong custodial dependence on
the normalization point. So, putting the normalization point $\mu$ in the
region of charmed quark mass, we find that the two-loop potential with the
three-loop running coupling constant $\alpha_s^{\overline{\Rsub MS}}$ has an
unremovable additive shift depending on $\mu$. This shift has variation in wide
limits. This fact illuminates the presence of infrared singularity in the
coupling constant of QCD, so that the $\mu$-dependent shift in the potential
energy has the form of a pole posed at $\Lambda_{QCD}$ \cite{3loop}.

Thus, in order to avoid the ambiguity of static potential in QCD we have to
deal with infrared stable quantities. The motivation by Buchm\"uller and Tye
was to write down the $\beta$ function of $\alpha_{\Rsub V}$ consistent with
two known asymptotic regimes at short and long distances. They proposed the
function, which results in the effective charge determined by two parameters,
only: the perturbative parameter is the scale in the running of coupling
constant at large virtualities and the nonperturbative parameter is the string
tension. The necessary inputs are the coefficients of $\beta$ function. The
parameters of potential by Buchm\"uller and Tye were fixed by fitting the mass
spectra of charmonium and bottomonium \cite{PDG}. Particularly, in such the
phenomenological approach the scale $\Lambda_{\overline{\Rsub
MS}}^{n_f=4}\approx 510$ MeV was determined. It determines the asymptotic
behaviour of coupling constant at large virtualities in QCD. This value is in a
deep contradiction with the current data on the QCD coupling constant
$\alpha_s^{\overline{\Rsub MS}}$ \cite{PDG}. In addition, one can easily find
that the three-loop coefficient $\beta_2^{\Rsub V}$ for the $\beta$ function
suggested by Buchm\"uller and Tye is not correct even by its sign and absolute
value in comparison with the exact coefficient recently calculated  in
\cite{Peter,Schroed}.

Thus, the modification of Buchm\"uller--Tye (BT) potential of static quarks as
dictated by the current status of perturbative calculations is of great
interest.

The nonperturbative behaviour of QCD forces between the static heavy quarks  at
long distances $r$ is usually represented by the linear potential (see
discussion in ref.\cite{simon})
\begin{equation}
V^{\rm conf}(r) = k\cdot r, \label{conf}
\end{equation}
which corresponds to the square-law limit for the Wilson loop. The form of
(\ref{conf}) corresponds to the limit, when at low virtualities ${\bf q}^2\to
0$ the coupling $\alpha_{\Rsub V}$ tends to
$$
\alpha_{\Rsub V}({\bf q^2}) \to \frac{K}{\bf q^2},
$$
so that
\begin{equation}
\frac{d \alpha_{\Rsub V}({\bf q^2})}{d \ln \bf q^2} \to - \alpha_{\Rsub V}({\bf
q^2}),
\label{lim-c}
\end{equation}
which gives the confinement asymptotics for the $\beta_{\Rsub V}$ function.

Buchm\"uller and Tye proposed the procedure for the reconstruction of
$\beta$ function in the whole region of charge variation by the known limits of
asymptotic freedom to a given order in $\alpha_s$ and confinement regime.
Generalizing their method, the $\beta_{\rm PT}$ function found in the framework
of asymptotic perturbative theory (PT) to three loops, is transformed to the
$\beta$ function of effective charge as follows:
\begin{eqnarray}
\displaystyle
\frac{1}{\beta_{\rm PT}({\mathfrak a})} &=& -\frac{1}{\beta_0 {\mathfrak a}^2}
+
\frac{\beta_1+\left(\beta_2^{\Rsub V} - \frac{\beta_1^2}{\beta_0}\right)
{\mathfrak a}}{\beta_0^2 {\mathfrak a}} \Longrightarrow \nonumber \\
\frac{1}{\beta({\mathfrak a})} &=& -\frac{1}{\beta_0 {\mathfrak a}^2 \left(1-
\exp\left[-\frac{1}{\beta_0 {\mathfrak a}}\right]\right)}+
\frac{\beta_1+\left(\beta_2^{\Rsub V} - \frac{\beta_1^2}{\beta_0}\right)
{\mathfrak a}}{\beta_0^2 {\mathfrak a}}
\exp\left[-\frac{l^2 {\mathfrak a}^2}{2}\right],
\label{KKO}
\end{eqnarray}
where the exponential factor in the second term contributes to the
next-to-next-to-leading order at ${\mathfrak a}\to 0$. This function has the
essential peculiarity at ${\mathfrak a}\to 0$, so that the expansion is the
asymptotic series in ${\mathfrak a}$. At ${\mathfrak a}\to \infty$ the $\beta$
function tends to the confinement limit represented in (\ref{lim-c}).

The construction of (\ref{KKO}) is based on the idea to remove the pole from
the coupling constant at a finite energy, but in contrast to the ``analytic''
approach developed in \cite{analit},  the ``smothing'' of peculiarity occurs in
the logarithmic derivative of charge related with the $\beta$ function, but in
the expression for the charge itself. Indeed, in the one-loop perturbation
theory we have got
$$
\frac{d\ln {\mathfrak a}}{d\ln \mu^2} = - \beta_0 {\mathfrak a} = -
\frac{1}{\ln
\frac{\mu^2}{\Lambda^2}},
$$
because of 
$$
{\mathfrak a} = \frac{1}{\beta_0 \ln \frac{\mu^2}{\Lambda^2}},
$$
and the pole can be cancelled in the logarithmic derivative itself, so
that\footnote{The presentation of power correction in the form of charge
exponential was similarly used in \cite{kp}.}
$$
\frac{d\ln {\mathfrak a}}{d\ln \mu^2} \Longrightarrow - \beta_0 {\mathfrak a}
\left(1-\frac{\Lambda^2}{\mu^2}\right) \approx  - \beta_0 {\mathfrak a} 
\left(1- \exp\left[-\frac{1}{\beta_0 {\mathfrak a}}\right]\right).
$$
As we see in the perturbative limit ${\mathfrak a}\to 0$ the deviation in the 
$\beta$ function is exponentially small. The generalization of this idea to two
loops was done by Buchm\"uller and Tye.

Remember that the one and two-loop static potentials matched with the linear
term of confinement lead to the contradiction with the value of QCD coupling
constant extracted at the scale of $Z$ boson mass if we fit the mass spectra of
heavy quarkonia in such potentials. The static potential in the three-loop
approximation results in the consistent value of QCD coupling constant at large
virtualities \cite{3loop}.

In the perturbative limit the usual solution for the running coupling constant
\begin{eqnarray}
{\mathfrak a}(\mu^2) = \frac{1}{\beta_0 \ln
\frac{\mu^2}{\Lambda^2}}&&\left[1 - 
\frac{\beta_1}{\beta_0^2}\frac{1}{\ln
\frac{\mu^2}{\Lambda^2}} 
\ln \ln \frac{\mu^2}{\Lambda^2} + \right.
\nonumber \\
&&
\left.
\frac{\beta_1^2}{\beta_0^4}\frac{1}{\ln^2
\frac{\mu^2}{\Lambda^2}} 
\left( \ln^2 \ln \frac{\mu^2}{\Lambda^2} - \ln \ln \frac{\mu^2}{\Lambda^2} -1 +
\frac{\beta_2^{\Rsub V} \beta_0}{\beta_1^2}\right)\right],
\label{3pt}
\end{eqnarray}
is valid. Using the asymptotic limit of (\ref{3pt}), one can get the equation
\begin{eqnarray}
\ln \frac{\mu^2}{\Lambda^2} & = & \frac{1}{\beta_0 {\mathfrak
a}(\mu^2)}+
\frac{\beta_1}{\beta_0^2} \ln \beta_0 {\mathfrak
a}(\mu^2)+\int_0^{{\mathfrak a}(\mu^2)} dx 
\left[\frac{1}{\beta_0 x^2}-\frac{\beta_1}{\beta_0^2
x}+\frac{1}{\beta(x)}\right], \label{ptlim} 
\end{eqnarray}
which can be easily integrated out, so that we get an implicit solution for the
charge depending on the scale. The implicit equation can be inverted by the
iteration procedure, so that well approximated solution has the form
\begin{equation}
{\mathfrak a}(\mu^2) = \frac{1}{\beta_0 \ln\left(1+\eta(\mu^2)
\frac{\mu^2}{\Lambda^2}\right)},
\label{eff}
\end{equation}
where $\eta(\mu^2)$ is expressed through the coefficients of perturbative
$\beta$ function and parameter $l$ in (\ref{KKO}), which is related to the
slope of Regge trajectories $\alpha^\prime_P$ and the integration constant, the
scale $\Lambda$ \cite{3loop}.

The slope of Regge trajectories, determining the
linear part of potential, is supposed equal to
$ 
\alpha^\prime_P = 1.04\;{\rm GeV}^{-2},
$ so that in (\ref{conf}) we put the parameter
$
k = \frac{1}{2 \pi \alpha^\prime_P}.
$
We use also the measured value of QCD coupling constant \cite{PDG} and pose
$$
\alpha_s^{\overline{\Rsub MS}}(m_Z^2) = 0.123,
$$
as the basic input of the potential. The transformation into the configuration
space was done numerically in \cite{3loop}, so that the potential is presented
in the form of file in the notebook format of MATHEMATICA system. 

The analysis of quark masses and mass spectra of heavy quarkonia results in the
following values ascribed to the potential approach \cite{3loop}:
\begin{equation}
m_c^{\Rsub V} = 1.468\;{\rm GeV,}\;\;\;
m_b^{\Rsub V} = 4.873\;{\rm GeV.}
\label{mcmb}
\end{equation}
Thus, the spectroscopic characteristics of systems composed of nonrelativistic
heavy quarks are determined in the approach with the static potential described
above.

\section{Numerical estimates}

In the potential approach  the values of wave functions for the $1S$ states in
the systems of $\bar b b$ and $\bar c c$ are equal to
\begin{equation}
{\sqrt{4\pi}}\Psi_{\bar b b}(0) = 2.513\; {\rm GeV}^{3/2},\;\;\;
{\sqrt{4\pi}}\Psi_{\bar c c}(0) = 0.895\; {\rm GeV}^{3/2},
\end{equation}
which are the renormalization group invariants by construction of Schr\"odinger
equation with the static potential. Remember that this potential includes the
contribution of linear part providing the confinement of quarks. The same
term makes the infrared stability of effective charge and all of the potential
in NRQCD.

The analysis of leptonic constants of $\Upsilon$ and $\psi$ with account of
Wilson coefficient ${\cal K}$, derived in the perturbative NRQCD, was done in 
\cite{3loop}. For $\Upsilon$ we found that the variation of hard scale in 
wide limits $\mu_{\rm hard} = (1-2) m_{b}$ led to the presence of a stable
point $\mu_{\rm fact}$, in which the result was weakly sensitive to the
variation of hard scale. The stability was observed at $\mu_{\rm fact}\approx
2.2-2.7$ GeV, i.e. at the inter-quark distances characteristic for the size of
$1S$ level in the system of $\bar b b$.

For the leptonic constant of charmonium $J/\psi$ the stability point under the
variation of $\mu_{\rm hard}$ was located at the reasonable values of soft
scale of factorization $\mu_{\rm fact}\approx 1.05-1.35$ GeV. The position of
stable point was significantly dependent of changes in $\mu_{\rm hard}$ close
to the mass of charmed quark.

However, in that consideration the stability of results under the variation of
scale $\mu_{\rm fact}$  for the perturbative calculations concerning the Wislon
coefficient in NRQCD was not observed.

\begin{figure}[th]
\setlength{\unitlength}{0.7mm}
\begin{center}
\begin{picture}(100,105)
\put(-5,5){\epsfxsize=7.7cm \epsfbox{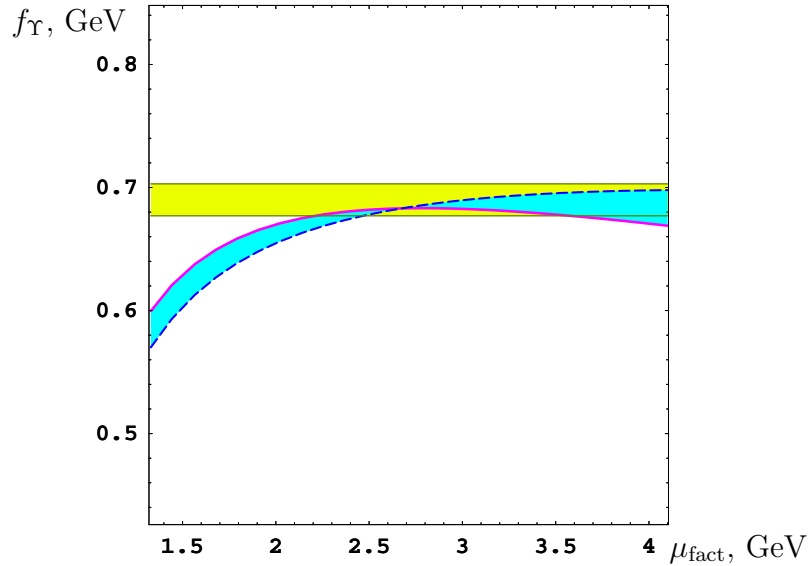}}
\put(105,8){$\mu_{\rm fact}$, GeV}
\put(-20,108){$f_{\Upsilon}$, GeV}
\end{picture}
\end{center}

\caption{The leptonic constant of ground vector state in the system of
bottomonium is presented versus the soft scale of normalization. The dashed
curve corresponds to $\mu_{\rm hard} = 2\, m_b$, while the solid curve does to
$\mu_{\rm hard} = m_b$. The initial condition for the evolution of
normalization factor ${\cal A}(\mu_{\rm fact})$ in the matrix element of
current in the nonrelativistic representation has been posed in the form ${\cal
A}(2.3\;{\rm GeV})=1$. The horizontal band is the experimental limits for the
constant.} 
\label{fups}
\end{figure}

In this paper we introduce the normalization factor ${\cal A}$ into the
analysis for the current of nonrelativistic quarks. As we see in Fig.
\ref{fups}, in the system of $\bar b b$ the choice of initial point of
normalization $\mu_0=2.3$ GeV in the mentioned region of stability for the
leptonic constant under the variation of matching scale, so that ${\cal
A}(2.3\;{\rm GeV})=1$, we get the independence of result for the lepton
constant of $\Upsilon$ in the interval $\mu_{\rm fact} = 1.8 - 3.8$ GeV.

\begin{figure}[ph]
\setlength{\unitlength}{0.7mm}
\begin{center}
\begin{picture}(115,90)
\put(-5,8){\epsfxsize=9cm \epsfbox{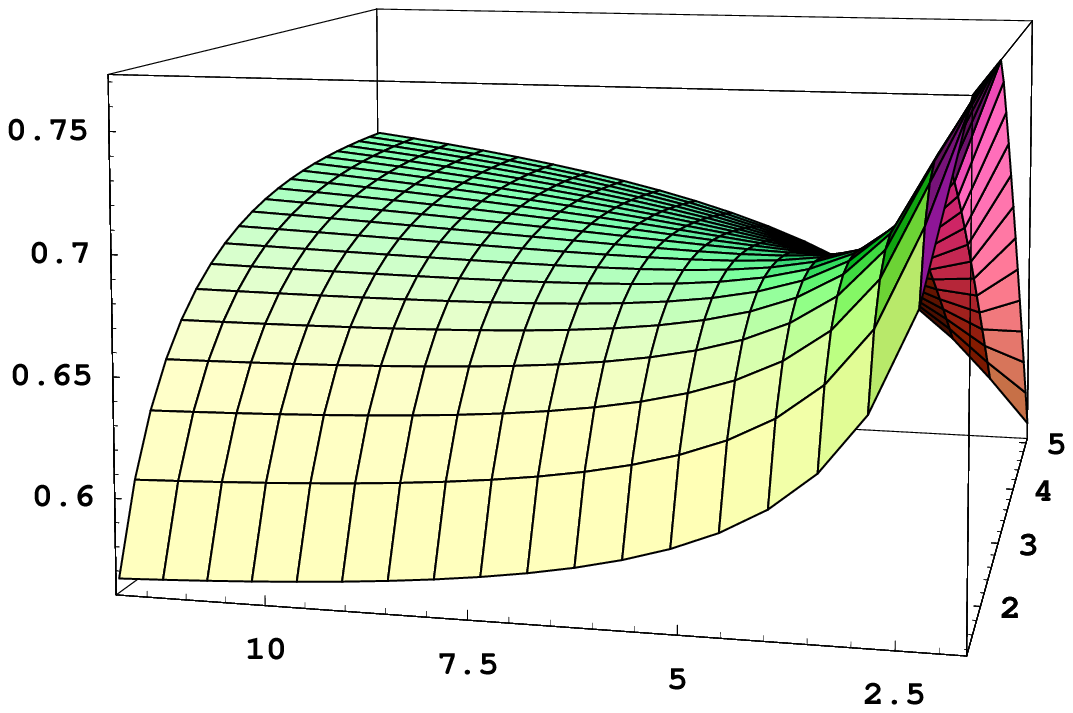}}
\put(125,25){$\mu_{\rm fact}$, GeV}
\put(45,9){$\mu_{\rm hard}$, GeV}
\put(-15,88){$f_{\Upsilon}$, GeV}
\end{picture}
\end{center}

\vspace*{-12mm}
\caption{The leptonic constant of ground vector state in the system of
bottomonium is presented versus the soft and hard scales, $\mu_{\rm fact}$ and
$\mu_{\rm hard}$. The initial condition for the evolution of
normalization factor ${\cal A}(\mu_{\rm fact})$ in the matrix element of
current in the nonrelativistic representation is ${\cal A}(2.3\;{\rm GeV})=1$.} 
\label{fups3d}
\end{figure}
\begin{figure}[ph]
\setlength{\unitlength}{0.75mm}
\begin{center}
\begin{picture}(100,100)
\put(-5,5){\epsfxsize=7cm \epsfbox{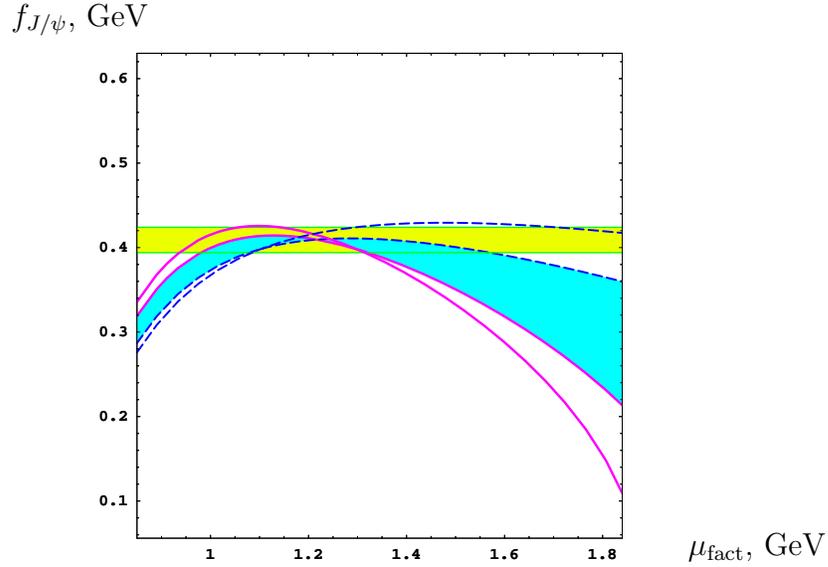}}
\put(100,8){$\mu_{\rm fact}$, GeV}
\put(-20,102){$f_{J/\psi}$, GeV}
\end{picture}
\end{center}

\vspace*{-12mm}
\caption{The leptonic constant of ground vector state in the system of
charmonium is presented versus the soft scale of normalization. The shaded
region restricted by curves corresponds to the change of hard scale from
$\mu_{\rm hard} = 2.0\, m_c$ (the dashed curve) to $\mu_{\rm hard} = 1.3\, m_c$
(the solid curve). The initial condition for the evolution of
normalization factor ${\cal A}(\mu_{\rm fact})$ in the matrix element of
current in the nonrelativistic representation has been posed in the form ${\cal
A}(1.07\;{\rm GeV})=1$. The horizontal band is the experimental limits for the
constant. The additional curves are presented for $\mu_{\rm hard} = 2.8\, m_c$
(the dashed line) and $\mu_{\rm hard} = 1.15\, m_c$ (the solid line).} 
\label{fpsi}
\end{figure}

The analysis of stability is performed in Fig. \ref{fups3d}. Thus, our estimate
of leptonic constant for the ground vector state of bottomonium is given by 
$$
f_{\Upsilon} = 685\pm 30\; {\rm MeV,}
$$
which is in a good agreement with the experimental data.

The calculations of leptonic constant for the charmonium $J/\psi$ are more
sensitive to the choice of initial condition in the evolution of factor ${\cal
A}$, because the stability point under the variation of hard scale for the
matching is significantly fluctuates under the change of $\mu_{\rm
hard}=(1-2)\, m_c$. In Fig. \ref{fpsi} we show the dependence of $f_{J/\psi}$
on the scale $\mu_{\rm fact}$ at ${\cal A}(1.07\;{\rm GeV})=1$. This value of 
$\mu_0=1.07$ GeV agrees with the inverse size of system $\bar c c$. We see in
the figure that the region of result stability is given by $\mu_{\rm fact} = 1.
- 1.6$ GeV.

\begin{figure}[th]
\setlength{\unitlength}{0.75mm}
\begin{center}
\begin{picture}(115,95)
\put(-5,8){\epsfxsize=9cm \epsfbox{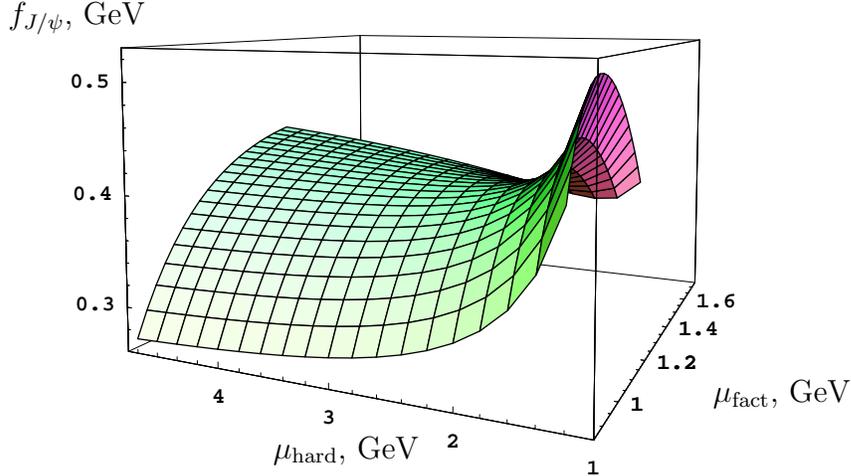}}
\put(110,25){$\mu_{\rm fact}$, GeV}
\put(32,15){$\mu_{\rm hard}$, GeV}
\put(-15,92){$f_{J/\psi}$, GeV}
\end{picture}
\end{center}

\vspace*{-12mm}
\caption{The leptonic constant of ground vector state in the system of
charmonium is presented versus the soft and hard scales, $\mu_{\rm fact}$ and
$\mu_{\rm hard}$. The initial condition for the evolution of
normalization factor ${\cal A}(\mu_{\rm fact})$ in the matrix element of
current in the nonrelativistic representation is given by ${\cal A}(1.07\;{\rm
GeV})=1$.} 
\label{fpsi3d}
\end{figure}

The dependence of estimates on the scales are presented in Fig. \ref{fpsi3d}.
Then we get
$$
f_{J/\psi} = 400\pm 45\; {\rm MeV,}
$$
which can be compared with the experimental value $f^{\rm exp}_{J/\psi} =
409\pm 15\; {\rm MeV}$.

Thus, the analysis of stability for the leptonic constants of heavy quarkonia
in the framework of potential approach shows that there is the region of
stability under the change of both factorization scale in the perturbative
NRQCD and point marking the matching of effective theory of nonrelativistic
quarks with the full QCD. The basic source of uncertainty in the estimates is
the variation of initial normalization point for the evolution factor
determining the matrix element for the current of nonrelativistic quarks, so
that the values of leptonic constants in the extremum (see Figs. \ref{fups3d}
and \ref{fpsi3d}) change in the limits, which are included into the methodic
error of estimates as shown above.

\begin{figure}[th]
\setlength{\unitlength}{1mm}
\begin{center}
\begin{picture}(100,66)
\put(5,5){\epsfxsize=8cm \epsfbox{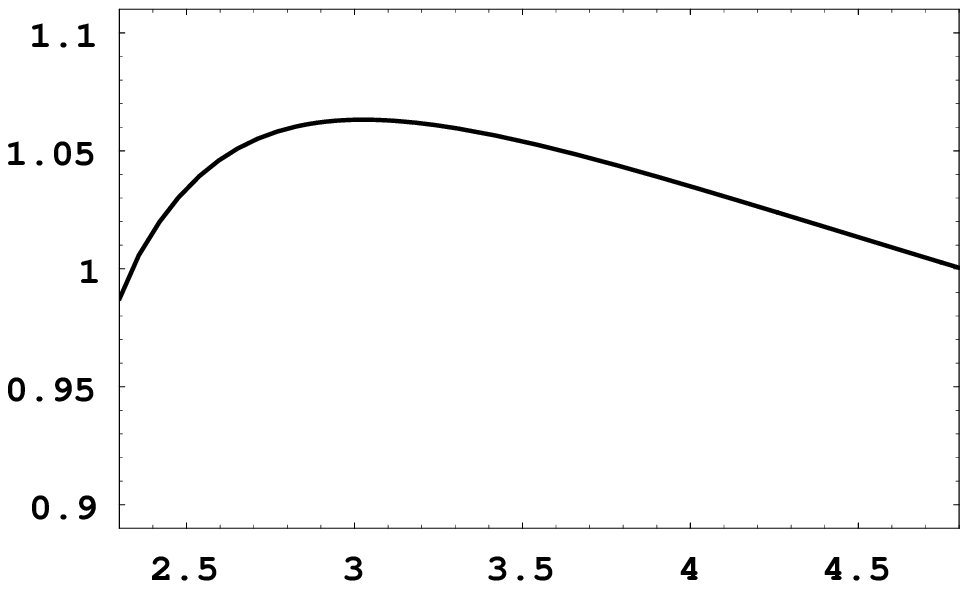}}
\put(85,3){$\mu=m_Q\cdot v$, GeV}
\put(5,60){$\displaystyle\frac{{\cal K}^{\rm vNRQCD}(v)}{{\cal K}^{\rm
vNRQCD}(1)}$}
\end{picture}
\end{center}
\caption{The Wilson coefficient ${\cal K}^{\rm vNRQCD}(v)$ for the vector
electromagnetic current of nonrelativistic bottom quarks in vNRQCD.}
\label{vNRQCD}
\end{figure}

In vNRQCD the factor ${\cal K}^{\rm vNRQCD}(v)$ demonstrates the stability
region (see Fig. \ref{vNRQCD} for the bottomonium). This stable point is the
only reasonable indication for the choice of $v_0$, at which ${\cal A}^{\rm
vNRQCD}(v_0)=1$. Such the prescription results in the estimates of leptonic
constants derived in vNRQCD
\begin{equation}
\left.f_\Upsilon\right|_{\rm vNRQCD} = 680\pm 30\; {\rm MeV,}\;\;\;
\left.f_{J/\psi}\right|_{\rm vNRQCD} = 360\pm 45\; {\rm MeV,}
\end{equation}
\noindent
where the errors are dominantly due to the variations of $\alpha_s$ at the
masses of bottom and charmed quarks: $\alpha_s(m_b) = 0.22\pm 0.02$,
$\alpha_s(m_c) = 0.34\pm 0.05$ entering the one-loop matching condition. We see
that the vNRQCD estimates are in agreement with both the static values and
experimental data.

Finally, in the potential approach we can calculate the ratios of leptonic
constants for the excited $nS$-wave levels in $\bar b b$ and $\bar c c$, which
were presented in \cite{3loop} in comparison with the experimental data. These
theoretical calculations are in a good agreement with the measured values.

\section{Conclusion}
In the present work we have done the general consideration of problem on the
calculation of leptonic constants for the heavy quarkonia in NRQCD and
introduced the factor for the normalization of matrix element for
the current of nonrelativistic quarks. This factor has got the anomalous
dimension of current in the effective theory, so that the explicit invariance
under the action of renormalization group takes place for the estimated values
of leptonic constants. In the framework of potential approach we have estimated
the constants for $f_\Upsilon$ and $f_{J/\psi}$, so that we have found
reasonable regions of stability under the variation of both the scale for the
matching of NRQCD with full QCD and the soft scale ascribed to the operators of
nonrelativistic quarks in NRQCD. The obtained results are in  a good agreement
with the experimental data.

The authors expresses their gratitude to V.A.Rubakov and S.A.Larin for the
stimulating remarks and the challenge, the response to which leads to the
writing of this work.

This work is in part supported by the Russian Foundation for Basic Research,
grants 01-02-99315, 01-02-16585 and 00-15-96645 and the Federal program ``State
support for the integration of high education and fundamental science'', grant
247 (V.V.K. and O.N.P.).


\newpage

\begin{thebibliography}{**}
\bibitem{SVZ}
{ M.A.Shifman, A.I.Vainshtein, V.I.Zakharov}, Nucl. Phys. {\bf B147}, 385
(1979);\\
{ L.J.Reinders, H.R.Rubinstein, S.Yazaki}, Phys. Rep. {\bf 127}, 1 (1985);\\
{ V.A.Novikov et al.}, Phys. Rep. {\bf 41C}, 1 (1978).
\bibitem{NRQCD}
G.T.Bodwin, E.Braaten, G.P.Lepage, Phys. Rev. {\bf D51}, 1125 (1995);\\
T.Mannel, G.A.Schuler, Z. Phys. {\bf C67}, 159 (1995).
\bibitem{PDG}
D.E.Groom et al., Eur. Phys. J. {\bf C15}, 1 (2000). 
\bibitem{mbv}
M.B.Voloshin, Int. J. Mod. Phys. {\bf A10}, 2865 (1995).
\bibitem{pp}
A.A.Penin, A.A.Pivovarov, Nucl. Phys. {\bf B549}, 217 (1999).
\bibitem{ben}
M.Beneke, hep-ph/9911490 (1999);\\
M.Beneke, A.Signer, Phys. Lett. {\bf B471}, 233 (1999).
\bibitem{hoang}
A.H.Hoang,  Phys. Rev. {\bf D61}, 034005 (2000).
\bibitem{eide}
M.Eidem\"uller, M.Jamin,  hep-ph/0010334 (2000).
\bibitem{3loop}
V.V.Kiselev, A.E.Kovalsky, A.I.Onishchenko, { hep-ph/0005020} (2000).
\bibitem{Peter}
M.Peter, Nucl. Phys. {\bf B501}, 471 (1997), Phys. Rev. Lett. {\bf 78}, 602
(1997).
\bibitem{Schroed}
Y.Schr\"oder, Phys. Lett. {\bf B447}, 321 (1999).
\bibitem{BT}
W.Buchm\" uller, S.-H.H.Tye, Phys. Rev. D24, 132 (1981).
\bibitem{HT}
A.H.Hoang, T.Teubner, Phys. Rev. {\bf D58}, 114023 (1998).
\bibitem{bensmir}
M.Beneke, A.Signer, V.A.Smirnov, Phys. Rev. Lett. {\bf 80}, 2535 (1998).
\bibitem{melch}
A.Czarnecki, K.Melnikov, Phys. Rev. Lett. {\bf 80}, 2531 (1998).
\bibitem{mel}
K.Melnikov, A.Yelkhovsky, Phys. Rev. {\bf D59}, 114009 (1999).
\bibitem{pNRQCD}
A.Pineda, J.Soto, Nucl. Phys. Proc. Suppl. {\bf 64}, 428 (1998).
\bibitem{LMR}
M.Luke, A.Manohar, I.Rothstein, Phys. Rev. {\bf D61}, 074025 (2000).
\bibitem{MS}
A.V.Manohar, I.W.Stewart,  Phys. Rev. {\bf D63}, 054004 (2001).
\bibitem{HMS}
A.Hoang, A.V.Manohar, I.W.Stewart, Phys. Rev. {\bf D64}, 014033 (2001).
\bibitem{Corn}
E.Eichten et al., Phys. Rev. {\bf D17}, 3090 (1979), {\bf D21}, 203 (1980).
\bibitem{log}
C.Quigg, J.L.Rosner, Phys. Lett. {\bf B71}, 153 (1977).
\bibitem{Mart}
A.Martin, Phys. Lett. {\bf B93}, 338 (1980).
\bibitem{RQ}
C.Quigg, J.L.Rosner, Phys. Rep. {\bf 56}, 167 (1979).
\bibitem{Richard}
J.L.Richardson, Phys. Lett. {\bf B82}, 272 (1979).
\bibitem{Su}
L.Susskind, {\em Coarse grained QCD} in R.~Balian and C.H.~Llewellyn Smith
(eds.), {\em Weak and electromagnetic interactions at high energy} 
(North Holland, Amsterdam, 1977);\\
W.Fischler, Nucl. Phys. {\bf B129}, 157 (1977);\\
T.Appelquist, M.Dine and I.J.Muzinich, Phys. Lett. {\bf B69}, 231 (1977), Phys.
Rev. {\bf D17}, 2074 (1978);\\
A.Billoire, Phys. Lett. {\bf B92}, 343 (1980);\\
E.Eichten and F.L.Feinberg, Phys. Rev. Lett. {\bf 43}, 1205 (1978), Phys. Rev.
{\bf D23}, 2724 (1981).
\bibitem{simon}
Yu.A.Simonov, Phys. Rept. {\bf 320}, 265 (1999),
Phys. Usp. {\bf 39}, 313 (1996);\\
Yu.A.Simonov, Uspekhi Fiz. Nauk {\bf 166}, 337 (1996);\\
Yu.A.Simonov, S.Titard, F.J.Yndurain, Phys. Lett. {\bf B354}, 435 (1995);\\
V.I.Zakharov, Phys. Rept. {\bf 320} 59 (1999);\\
M.N.Chernodub, F.V.Gubarev, M.I.Polikarpov, V.I.Zakharov, Phys. Lett. {\bf
B475}, 303 (2000);\\
K.G.Chetyrkin, S.Narison, V.I.Zakharov, Nucl. Phys. {\bf B550}, 353 (1999).
\bibitem{analit}
I.L.Solovtsov, D.V.Shirkov, Theor. Math. Phys. {\bf
120}, 1220 (1999);\\
I.L.Solovtsov, D.V.Shirkov, Phys. Rev. Lett. {\bf 79}, 1209 (1997), Phys. Lett.
{\bf B442}, 344 (1998);\\
A.I.Alekseev, hep-ph/9906304 (1999).
\bibitem{kp}
N.V.Krasnikov, A.A.Pivovarov, hep-ph/9510207 (1995).
%
%
%
\end{thebibliography}
\end{document}